\documentclass[%
 aip,
 amsmath,amssymb,
 reprint,%
]{revtex4-2}

\usepackage{graphicx}
\usepackage{dcolumn}
\usepackage{tikz}
\usepackage{bm}
\usepackage[utf8]{inputenc}
\usepackage[T1]{fontenc}
\usepackage{mathptmx}
\usepackage{etoolbox}

\usepackage[]{siunitx}
\usepackage{acronym}

\makeatletter
\def\@email#1#2{%
 \endgroup
 \patchcmd{\titleblock@produce}
  {\frontmatter@RRAPformat}
  {\frontmatter@RRAPformat{\produce@RRAP{*#1\href{mailto:#2}{#2}}}\frontmatter@RRAPformat}
  {}{}
}%
\makeatother

\begin{document}

\preprint{AIP/xxx-QED}

\title{Phonon dephasing times determined with time-delayed, broadband CARS}
\author{F. Hempel}
\affiliation{TU Dresden, Institute of Applied Physics, Nöthnitzer Strasse 61, 01187 Dresden, Germany}

\author{M. Rüsing}
\affiliation{Department of Physics, Paderborn University, Warburger Str. 100, 33098 Paderborn, Germany}
\affiliation{Institute for Photonic Quantum Systems (PhoQS), Paderborn University, Warburger Str. 100, 33098 Paderborn, Germany}

\author{F. Vernuccio}
\affiliation{Physics Department, Politecnico di Milano, P.zza L. da Vinci 32, 20133 Milano, Italy}

\author{K. J. Spychala}
\affiliation{Department of Physics, Paderborn University, Warburger Str. 100, 33098 Paderborn, Germany}

\author{R. Buschbeck}
\affiliation{TU Dresden, Institute of Applied Physics, Nöthnitzer Strasse 61, 01187 Dresden, Germany}

\author{G. Cerullo}
\affiliation{Physics Department, Politecnico di Milano, P.zza L. da Vinci 32, 20133 Milano, Italy}
\affiliation{CNR Institute for Photonics and Nanotechnology (CNR-IFN), P.zza L. da Vinci 32, 20133 Milano, Italy}

\author{D. Polli}
\affiliation{Physics Department, Politecnico di Milano, P.zza L. da Vinci 32, 20133 Milano, Italy}
\affiliation{CNR Institute for Photonics and Nanotechnology (CNR-IFN), P.zza L. da Vinci 32, 20133 Milano, Italy}

\author{L. M. Eng}
\affiliation{TU Dresden, Institute of Applied Physics, Nöthnitzer Strasse 61, 01187 Dresden, Germany}
\affiliation{ct.qmat: Dresden-Würzburg Cluster of Excellence—EXC 2147, TU Dresden, 01062 Dresden, Germany}

\email{lukas.eng@tu-dresden.de}
\email{dario.polli@polimi.it}

\date{\today}

\begin{abstract}
Coherent Raman scattering techniques as coherent anti-Stokes Raman scattering (CARS), offer significant advantages in terms of pixel dwell times and speed as compared to spontaneous Raman scattering for investigations of crystalline materials. However, the spectral information in CARS is often hampered by the presence of a non-resonant contribution to the scattering process that shifts and distorts the Raman peaks. In this work, we apply a method to obtain non-resonant background-free spectra based on time-delayed, broadband CARS (TD-BCARS) using an intra-pulse excitation scheme. In particular, this method can measure the phononic dephasing times across the full phonon spectrum at once. We test the methodology on amorphous SiO$_2$ (glass), which is used to characterize the setup-specific and material-independent response times, and then apply TD-BCARS to the analysis of single crystals of diamond and ferroelectrics of potassium titanyl phosphate (KTP) and potassium titanyl arsenate (KTA). For diamond, we determine a dephasing time of $\tau = 7.81$~ps for the single sp$^3$ peak.
\end{abstract}

\maketitle

\section{Introduction}

Spontaneous Raman (SR) micro-spectroscopy is extensively utilized in solid-state physics because the vibrational spectrum provides valuable information about numerous properties, such as crystal structure and its symmetry, the stoichiometric composition, presence of defects and dopants, ferroelectric and magnetic properties, local strain, or insights about nanoscale confinement, e.g. in nanoparticles or 2D materials \cite{Nataf2016,Fontana2015,Tejarina2014,Tejarina2013,Stone2011,Arora2007,Sunny2021,Sun2021,Paillet2018,Singh2023,Ruesing2018}. SR micro-spectroscopy requires little sample preparation and can be employed for 2D and even 3D mapping, when used in a confocal setup. However, a key challenge of SR is its weak scattering cross-section, which often results in long integration times of several seconds or more per spectrum, severely limiting acquisition speed and, in turn, the field of view in image applications \cite{Parodi2020,Zhang2021CARS,Li2021}. In this regard, coherent anti-Stokes Raman scattering (CARS) offers a viable alternative\cite{Parodi2020,Zhang2021CARS,Li2021,Reitzig2021}. CARS sees widespread use in microbiology for chemically-sensitive detectiong and imaging and has recently also been applied to solid-state crystal systems. CARS is based on a nonlinear optical scattering process, which enables integration times per spectrum in the \textmu s regime and below, which enables high-speed single-frequency vibrational imaging over large fields of view. This is particularly true for broadband CARS (BCARS) methods \cite{Hempel2021,Hempel2023ComparingTA,Vernuccio2022FingerprintMC,Vernuccio2024,Malard2021,Xu2022}, which probe a full spectrum at once without the need for lengthy tuning of one or more of the involved lasers. In principle, CARS offers comparable information to SR spectroscopy, because it also provides spectral information about the phonons of a material, their frequencies and scattering efficiency, but with much shorter integration times.

Nonetheless, CARS comes with challenges, such as the presence of a non-resonant background (NRB), which leads to distortion of the spectral information, e.g. a spectral shift, as well as asymmetric broadening of peaks \cite{Hempel2021,Hempel2023ComparingTA,Vernuccio2024,Camp2016,Muddiman2023,Valensise2020}. The NRB, which arises from the (broadband) electronic contribution to the nonlinear susceptibility, interferes with the resonant vibrational response and distorts the CARS spectra. This is a particular challenge for spectroscopic investigations, such as in the context of crystalline materials. Here, the information, e.g., about strain, defects, etc., is often encoded in subtle spectral changes, such as small ($<$1~cm$^{-1}$) shifts or the (asymmetric) broadening of peaks. Therefore, to accurately extract spectral information and achieve comparability to established SR or infrared (IR) spectroscopy measurements, it is mandatory to remove the NRB and extract the pure resonant response from the CARS spectra.

Consequently, various methods have been developed to retrieve NRB-free CARS spectra. Traditionally, the recorded raw-data spectra are corrected post-measurement by using numerical phase-retrieval algorithms based on either the time-domain Kramers-Kronig  \cite{Camp2016} or the maximum entropy methods \cite{Rinia2007}, which require a measurement on non-resonant samples (e.g., glass coverslip or water) to estimate the actual NRB of the material, as well as the setup-specific transfer function. However, since the pure non-resonant CARS response of the sample cannot be measured independently, the algorithms have to account for amplitude and phase differences that need to be corrected for prior to deriving the pure resonant response. Recently, the NRB removal problem has been treated as a supervised learning problem by using deep neural networks, which provides another valuable alternative\cite{Vernuccio2024,Muddiman2023,Valensise2020} for NRB removal. These methods do not need any additional measurements on reference samples; moreover, they provide the pure vibrational spectrum in a much shorter time than the one required by NRB removal algorithms, being beneficial for CARS imaging applications. 

\begin{figure*}[hbt!]
	\centering
	\includegraphics[width=0.8\linewidth]{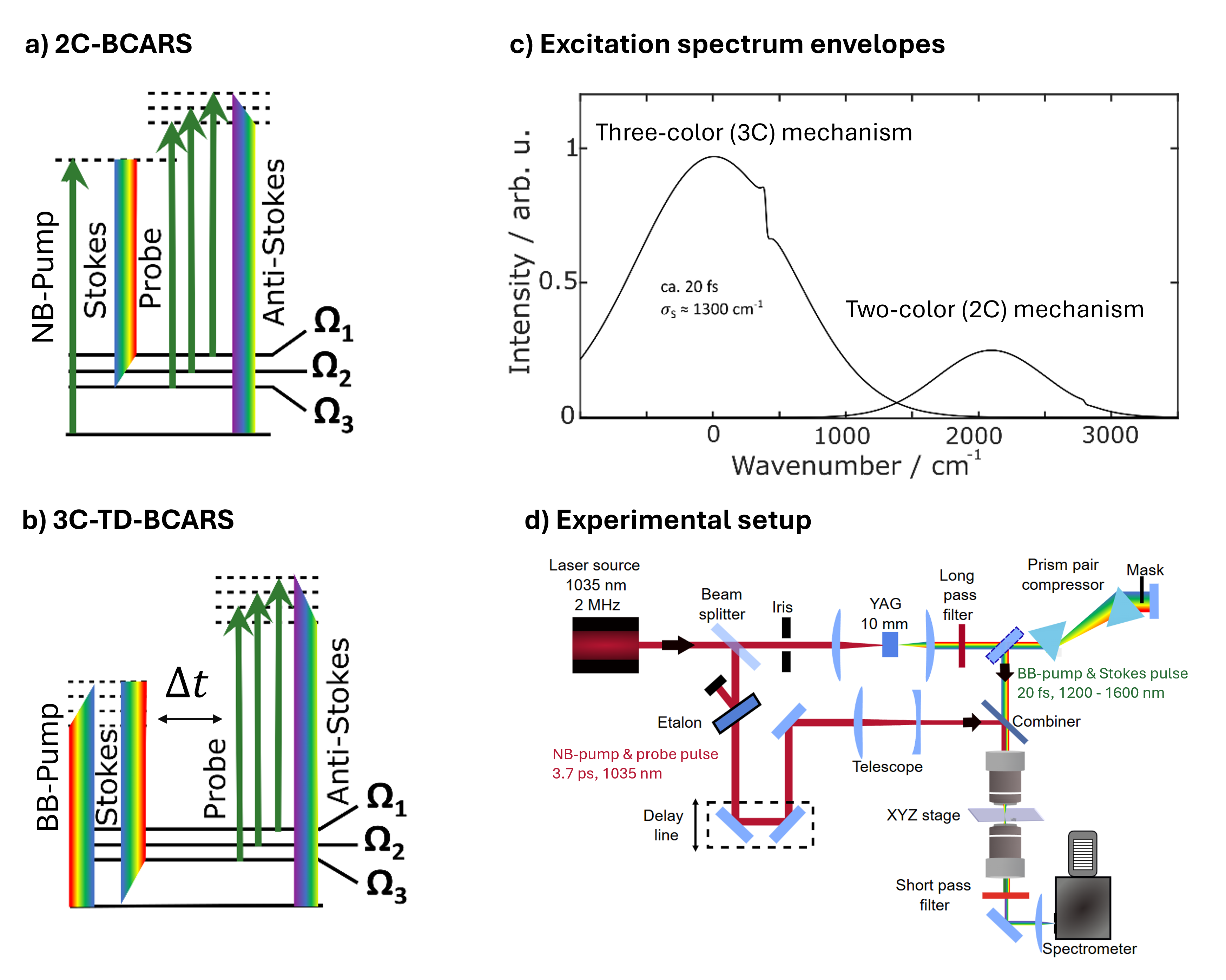}
	\caption[TD-CARS Jablonski Diagram]{Jablonski diagrams for a) 2C-BCARS and b) 3C-TD-BCARS schemes. Please note the time-delay $\Delta t$ that can be introduced between the intra-pulse excitation and the probe pulse in the 3C-TD-BCARS setup. The key distinction between the 2C- and 3C-schemes are the uses of a narrowband (NB)-pump pulse in the 2C-scheme, and a broadband (BB) pump pulse in the 3C-scheme.  c) Excitation envelope for a sub-20~fs Stokes pulse centered around 1400~nm and a pump laser at 1035~nm. Here, phonons with wavenumbers of up to 1300~cm$^{-1}$ can be probed in the 3C-scheme, while 2C-spectrum is mostly separated from the 3C-region. Using an even lower probe laser wavelength or longer central wavelength for the Stokes pulse could separate the two regions even further. d) Sketch of the experimental setup.}
	\label{fig:TD_jablonski}
\end{figure*}

In this work, we present an experimental implementation based on time-delayed CARS (TD-CARS) to remove the NRB in the CARS spectra without the need of any post-processing algorithm. TD-CARS\cite{Pestove2007,Selm2010} is based on the fact that the non-resonant and resonant contributions in the CARS four-wave mixing process decay with substantially different time constants. Here, the non-resonant contribution arises from the response of the material's electronic system, whose dephasing is effectively instantaneous for the purposes of the present study\cite{Ogilvie2008TimedelayedCR}. In contrast, the resonant contribution in CARS is a response of the much slower moving atomic contributions (phonons) to the nonlinear susceptibility, which has dephasing times in the order of ps or longer, which is inversely proportional to the linewidth. Therefore, if a sufficient time-delay is introduced between the pump and probe pulses in the CARS process, a true NRB-free CARS spectrum is obtained.

Time-delayed CARS (TD-CARS), where a delay is introduced between pump and probe pulses, is a well-established method in biology and chemistry \cite{Sylvester2022,Ariunbold2020,Neethling2023,Choi2023,Pestov2007,Ariunbold2017} and has recently been extended to solid-state systems \cite{Koivistoinen2017}. Beyond suppressing the NRB, TD-CARS provides access to vibrational dephasing times, offering valuable information on material quality and properties. A major challenge, however, is that many TD-CARS implementations rely on laser tuning for hyperspectral analysis, or employ spectral bandwidths not optimized for the analysis of crystalline materials.

Instead, our experimental implementation combines broadband CARS and time-delayed CARS (TD-BCARS) to measure NRB-free spectra across a broad Raman range up to approximately $1300$~cm$^{-1}$, which is sufficient for most crystalline materials, which rarely have phonon frequencies higher than 1000~cm$^{-1}$. We achieve this by using an intra-pulse excitation scheme with a pulse of suitable bandwidth. We first characterize the NRB-dephasing times and their relation to the pump and probe pulses, before demonstrating NRB-free BCARS spectra. Furthermore, we use a controlled time-delay to determine phonon dephasing times, i.e. the lifetimes of the coherently excited vibrational modes, of crystalline materials across the full spectrum up to approximately 1300 cm$^{-1}$ at once. This latter information might be particularly interesting for research on solid-state materials, because the lifetimes of the coherently excited vibrations can be expected to be correlated, for example, with crystal quality or local material changes, e.g. by doping, etching, or other processes.

\begin{figure}[h]
 	\centering
 	\includegraphics[width=1\linewidth]{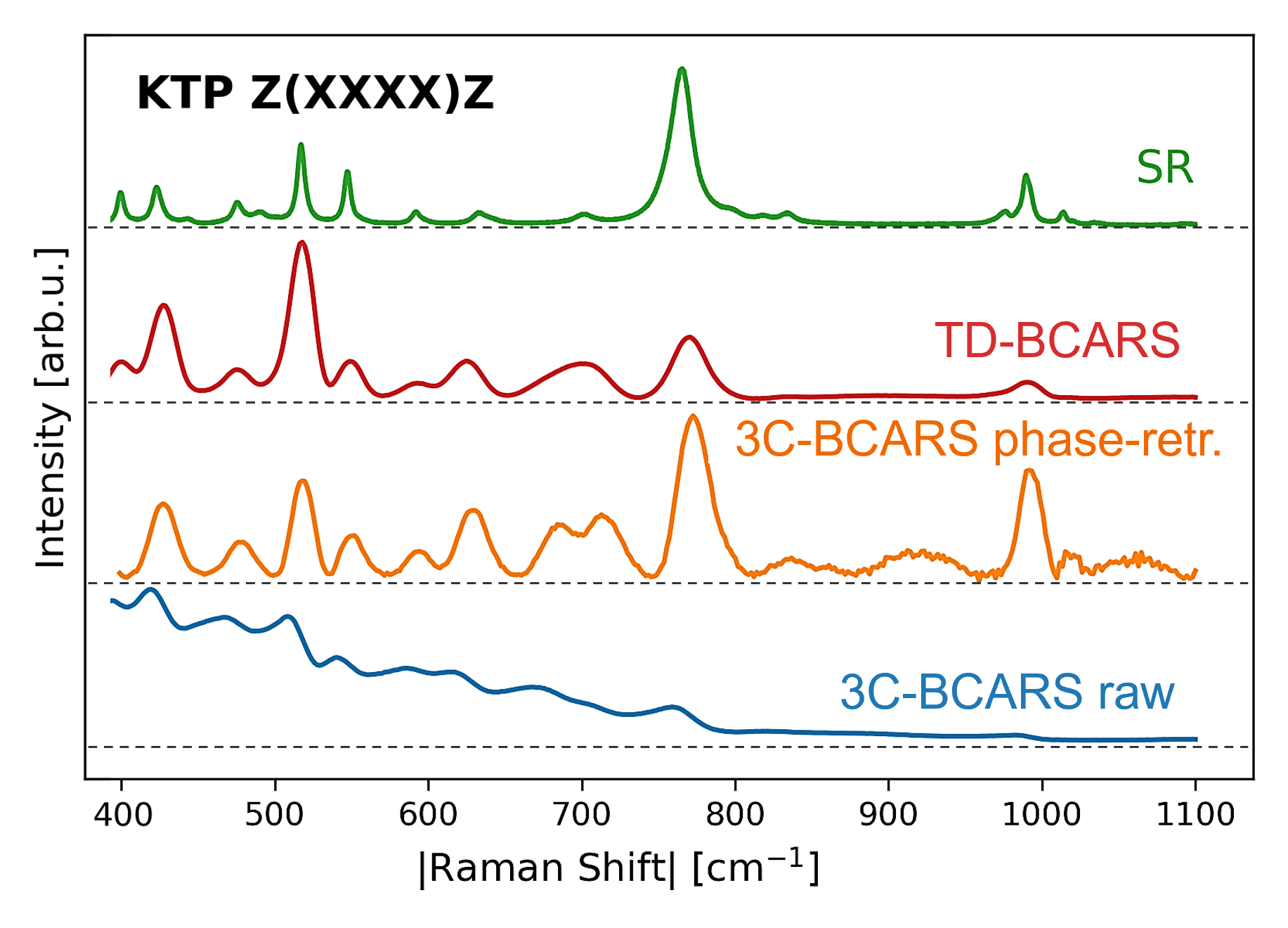}
 	\caption[TD-CARS and 3C-CARS Spectra of KTP]{The zero-time delay 3C-BCARS spectrum (blue) on KTP displays peaks down to \SI{400}{\per\cm}, necessitating the removal of the NRB contribution  for peak shape restoration by perforimg a phase-retrieval in-post processing \cite{Camp2016} (orange, 3C-BCARS phase-retr.). The TD-CARS measurement with a delay of 4~ps (red) demonstrates an NRB-free spectrum, eliminating the need for any post-processing. For reference, a SR spectrum of the same crystal is shown (green). Please note, the CARS spectra show a broader peak width of the phonon lines compared to the SR spectrum. This is a result of the spectral bandwidth of the probe laser, which is a 3.7~ps pulse at 1035 nm (approximately 10~cm$^{-1}$ bandwidth). This bandwidth is convoluted with the intrinsic line-width of the phonons. A higher resolution is possible if longer probe pulses are used, see F. Hempel et al., for example \cite{Hempel2023ComparingTA}. However, using spectrally narrow and longer probe pulses will inevitably limit the temporal resolution for TD-CARS. The Raman shift is given in absolute values to make the anti-Stokes BCARS signal more readily comparable with SR spectra, which are often Stokes-shifted.}
 	\label{fig:TD_CARS_spectra}
 \end{figure}

\section{The principles of time-delayed CARS}

CARS is a third-order nonlinear ($\chi^{(3)}$) four-wave mixing process involving three interactions with fields at the pump frequency $\omega_{pu}$, the Stokes frequency $\omega_{s}$, and the probe frequency $\omega_{pr}$, resulting in the emission of a signal at the anti-Stokes frequency $\omega_{as} = \omega_{pu}-\omega_{s}+\omega_{pr}$. In principle, CARS can use three distinct laser wavelengths for excitation. The intensity of the emitted anti-Stokes frequency light $\omega_{as}$ is drastically enhanced compared to non-resonant frequency combinations, whenever the difference between pump and Stokes frequency $\omega_{pu}-\omega_s = \Omega$ matches a vibrational frequency of a phonon mode $\Omega$, due to resonances in the frequency-dependent third-order nonlinear susceptibility $\chi^{(3)}$. Often, the pump and probe frequencies are chosen to be provided by the same laser, i.e. $\omega_{pr} = \omega_{pu}$, which also ensures temporal overlap between these beams. As only one separate Stokes frequency $\omega_s$ is required, this process is referred to as two-color CARS (2C-CARS) spectroscopy and is schematically shown in Fig.~\ref{fig:TD_jablonski}a). For that case the pump laser usually features a comparatively narrowband (NB-pump) spectrum. This is the most common procedure to excite and probe the vibrational modes in CARS implementations. In the case of BCARS, a broadband Stokes spectrum is used to generate broadband anti-Stokes components that contain the information from many vibrational modes at once\cite{Polli2018}. However, in these implementations, as pump and probe frequencies are provided by the same laser, the excitation and probing occurs instantaneously and cannot be temporally separated. Therefore, TD-CARS is not possible with a 2C-(B)CARS setup.

To allow for time-delayed CARS in broadband implementations, a three-color CARS (3C-CARS) excitation scheme is necessary as shown in Fig.~\ref{fig:TD_jablonski}b). Here, the generation of the excited vibrational states is provided by intra-pulse excitation with an ultra-short and spectrally broadband pump (BB-pump) pulse. Intra-pulse excitation is possible, if the Stokes pulse is shorter than the oscillation period for phonons (e.g., 24~fs for 1380~cm$^{-1}$), entailing that sub-20-fs pulses are needed to excite the vibrational modes up to $1300$ cm$^{-1}$. Such a short pulse also has the advantage of being spectrally broad. This means that intra-pulse excitation over a broad frequency range is possible. For typical crystalline materials, frequencies of only up to 1000~cm$^{-1}$ are required. To satisfy this requirement, in our experimental setup we generate  sub-20-fs pulses centered at 1400~nm, which are then combined with narrowband probe pulses at 1035~nm. The spectral coverage of our pulse is shown in Fig.~\ref{fig:TD_jablonski}c). Another advantage of the 3C-CARS process for the investigation of solid-state crystals is the possibility of exciting low-frequency vibrational modes, where a large number of intra-pulse combinations of wavelengths is possible. Also note that a 2C-CARS process similar to Fig.~\ref{fig:TD_jablonski}a) is inherently present in the 3C-CARS setup as well. In the 2C-CARS process, the sub-20-fs laser acts as the broadband Stokes pulse, while the probe laser from the 3C-scheme will provide the narrowband (NB) pump and probe pulses in the 2C-scheme. However, the 3C- and 2C-CARS processes do not cover the same wavelength range at the detector since they are used to probe different regions of the vibrational spectrum. The 3C-process probes the vibrational modes from 0 up to 1400 cm$^{-1}$, while the 2C-regime probes the vibrational modes from 1400 cm$^{-1}$ up to 3200 cm$^{-1}$. Larger spectral separation is possible by reducing the bandwidth of the broadband beam. A time-delayed operation can now readily be realized in the 3C-scheme by the introduction of a time delay between the broadband excitation and the probe pulse. As shown below, the 3C- and 2C-CARS regions are also distinguished by different temporal behaviors. Here, the 2C-regime can be fully suppressed when a sufficient time delay is introduced, while for the 3C region, first the NRB is reduced, while resonantly excited phonons can still be measured if their dephasing time is long enough. This allows to measure NRB-free spectra in the 3C-regime. Please note, that for larger time delays also the resonant contributions will eventually disappear due to their limited life-times. Therefore, a balance needs to be found, where the NRB is suppressed, while sufficiently large resonant signals are measurable.

The experimental setup is shown in Fig.~\ref{fig:TD_jablonski}d). The original laser source for both the narrowband and broadband pulses is provided by a Ytterbium laser operating at 1035~nm with a 2~MHz repetition rate. Compared to more conventional 80 MHz pump lasers, this system provides a higher energy per pulse. The laser provides 270~fs pulses, with 5~W average power. The fundamental laser beam is split into two arms. The top arm with 3~W, corresponding to 1.5~$\mu$J pulse energy, is used to generate the broadband pulse via supercontinuum generation in a YAG crystal (Yttrium aluminium granate).  This beam is sent into a pulse compressor to create the ultra-short 20~fs pulses necessary for the 3C-CARS intra-pulse excitation scheme. A mask is placed after the second prism of the compressor to narrow the pulse bandwidth in the range between 1200 and 1600 nm. After the compressor, the broadband beam features 50~mW average power. The bottom arm  with 2~W average power generates narrowband pulses through a high-finesse Fabry-Perot etalon,  yielding 3.7~ps pulses. The narrowband beam acts as a probe beam in our TD-CARS system and is responsible for the spectral resolution in CARS spectra, which is below \SI{10}{\per\cm}.  In the second arm, a mechanical delay stage is used to vary the delay time, $\Delta t$, between the pump/probe pulses. The delay line is controlled by a micrometer screw with 10 $\mu$m increments, which translates to a 67~fs accuracy of $\Delta t$. The two beams are combined through a dichroic mirror and sent to a homebuilt transmission microscope in up-right configuration equipped with two identical 100x air objectives with NA = 0.85.  The signal is collected in the forward direction and sent to a spectrometer, after rejection of the original beams by mean of a short-pass filter (FESH1000, Thorlabs). The sample can be raster-scanned in three dimensions using a motorized XYZ translation stage synchronized with the CCD camera of the spectrometer. However, in this work only point spectra are taken. More details on the setup can be found in previous works\cite{Vernuccio2022FingerprintMC,Hempel2023ComparingTA}.

To demonstrate the viability of NRB-suppression in the 3C-CARS configuration, measurements were conducted on a potassium titanyl phosphate (KTP) crystal using a Z(XXXX)Z (forward-) scattering configuration, with the results being depicted in Figure~\ref{fig:TD_CARS_spectra}. The scattering configuration is given in an extended Porto notation \cite{Hempel2021}, where the outer two Z denote the vectorial direction of the excitation and scattered light in crystal axis, respectively, while the inner X denote the polarization of the anti-Stokes signal, the probe, pump and Stokes beams, respectively. In this experiment, the light is focused parallel to the z-axis into the crystal and collected in forward direction, while the polarization of all lasers and the signal are chosen to be parallel to the x-axis of the crystal.

The unprocessed 3C-BCARS data with zero time-delay (blue) showcases the pronounced intensity for low-frequency modes down to \SI{400}{\per\cm}. This increasing intensity towards lower wavenumbers is a direct result of the intra-pulse excitation scheme, which provides more and more combinations of excitation wavelengths for smaller and smaller shifts. In principle, even lower-frequency modes are populated in the intra-pulse scheme, however, due to the filters used in the detection path of the experimental setup, a cut-off at $\sim \SI{400}{\per\cm}$ is observed. Note that, in principle Raman peaks down to a few \SI{10}{\per\cm} are measurable with such a setup, which is ideal for solid-state investigations. This spectrum is then subject to NRB removal using the Kramers-Kronig (KK) transformation \cite{Camp2016} which is applied to unveil the phase-retrieved CARS spectrum (orange, 3C-BCARS phase-retrieved).
In contrast, by choosing a time delay of approximately 4~ps, an NRB-free CARS spectrum can be obtained with TD-BCARS (red), eliminating the necessity for NRB removal in post processing. 
Notably, the relative peak intensities differ between TD-BCARS and the transformed 3C-BCARS. This disparity arises from the fact that in 3C-BCARS, the maximum intensity of each peak is measured, while in TD-CARS, the intensities  of the individual modes are reduced depending on their lifetimes. Further investigation into these lifetimes is presented below. Nevertheless, a spectrum with Lorentzian-like shapes similar to a typical SR spectrum (green) is readily obtained without the need for any processing.

TD-CARS proves to be an effective technique for experimental NRB removal and can validate the NRB removal by post-processing, for example via using the KK-algorithm. However, we must note that introducing a time delay also diminishes the intensity of the resonant component. In this case, the TD-CARS signal intensity is only \SI{10}{\percent} of the 3C-CARS peak intensity for zero time-delay, necessitating longer acquisition times. To combine the advantages of both techniques, 3C-BCARS without time-delay could be best suited for comprehensive mapping with subsequent KK transformation, while TD-BCARS excels in single-spectrum acquisition at specific points of interest, to validate the transformation results, as well as to extract phononic dephasing times, which will be discussed below.

\section{Temporal profile of the NRB}

\begin{figure*}[htb]
	\centering
	\includegraphics[width=0.750\textwidth]{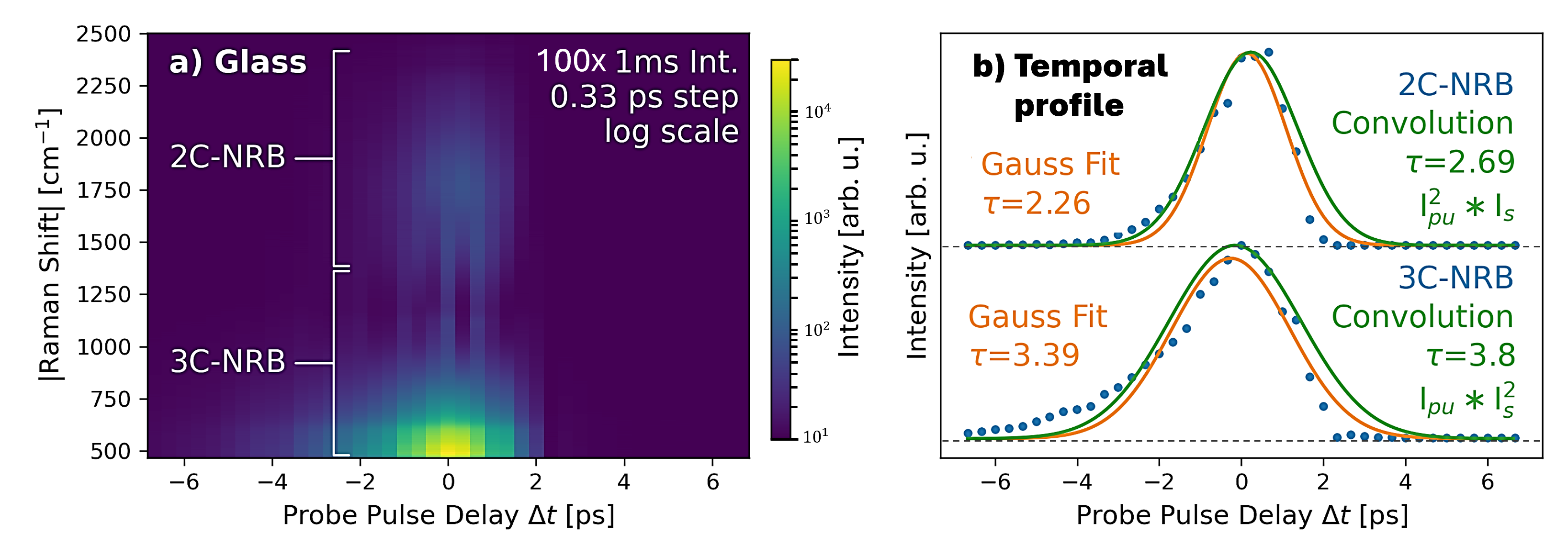}
	\caption[Time-Delay Sweep Glass]{TD-BCARS behavior of the NRB measured on glass: a) The spectral response shows two destinct regions, a 3C-NRB for wavenumbers approximately smaller than <\SI{1300}{\per\cm} and a 2C-NRB spanning from approximately 1300 to \SI{2500}{\per\cm}. b) The time-dependent intensity curves can be approximated by the convolution of the \SI{20}{\femto\s} Stokes pulse $I_s$ and the \SI{3.8}{\ps} pump pulse $I_{pu}$, which enters the process quadratically for the 2C-process, but only linearly in the 3C-process.}
	\label{fig:TD_NRB}
\end{figure*}

To enable an NRB-free measurement as well as to determine phonon dephasing times, it is first necessary to determine the duration of the NRB, which is mostly limited by the temporal convolution of the pulses with the instantaneous response of the NRB. For this, glass is investigated, which is an amorphous material with a frequency-independent third-order nonlinear response in the spectral range of interest \cite{Hempel2021,Hempel2023ComparingTA}. Glass shows no distinct phonon peaks due to its nature as an amorphous substance, but only an NRB. The time-delay sweeps are displayed in Figure~\ref{fig:TD_NRB}.

Fig.~\ref{fig:TD_NRB}~a) shows a hyperspectral intensity map plotted with the delay time on the x-axis and the Raman shift along the y-axis. The graph displays two distinct intensity regions: the 3C-NRB in the region on the bottom up to ca. \SI{1300}{\per\cm}, where intra-pulse excitation dominates, and the 2C-NRB region beyond approximately \SI{1300}{\per\cm}, where the narrowband probe laser also acts as as the pumplaser, while the 20-fs pulse acts as the broadband Stokes pulse. Fig.~\ref{fig:TD_NRB}~b) shows exemplary intensity curves for the two regions at approximately \SI{800}{\per\cm} (3C-NRB) and \SI{1700}{\per\cm} (2C-NRB). Both 2C- and 3C-signals display a somewhat a bell-shaped curve, with $\Delta t = 0$ being chosen from the position of maximum intensity. Basically, the observed shape and peak is similar to a form of correlation function of the (temporal) overlap of the Stokes and pump pulses. Here, a Gaussian fit of the 2C-NRB yields a temporal FWHM of about $\tau = \SI{2.26}{\pico\second}$, while the 3C-NRB yields a width of $\tau = \SI{3.39}{\pico\second}$. The Gaussian fits only approximately reproduce the measured curve: A slower buildup is observed for negative delays while the decay at positive delays is considerably faster. This asymmetry is most likely caused by pulse dispersion in the glass sample or by slight differences in the spectral/temporal shape of one or both pulses. Despite this, the experimental data already indicates that the nonresonant background (NRB) has largely decayed by delays greater than $\approx 3$~ps.

To understand the difference of almost 1~ps between the 2C- and 3C-NRB decay times, the temporal shapes can be estimated from the pulse durations of the Stokes and pump pulses and their interaction, respectively. Here, the NRB is expected to decay instantaneous. Therefore, the measurement is mainly broadened by the temporal width of the excitation pulse~\cite{Ogilvie2008TimedelayedCR}, which can be calculated as the cross-correlation of the pump, Stokes and probe pulses. For a simple estimation, both pulses are approximated with a Gaussian shape in the temporal domain. Due to the intra-pulse excitation, the 3C intensity $I_{3C}$ scales quadratically with the Stokes intensity $I_s$ of the broadband Stokes pulse, which in this case acts as both BB pump and Stokes, and linearly with the probe pulse intensity $I_{pr}$. Hence, the convolution is calculated as:

\begin{align}
    I_{3C}(\Delta t) \propto (I_{pr} \ast I_s^2)(\Delta t).
\end{align}

Here, convolution of the broadband Stokes pulse (estimated to be about 20~fs) and the 3.70~ps long pump pulse yields a convoluted duration of about $\tau_c = \SI{3.79}{\pico\second}$, which is reasonably consistent with the Gaussian fit of the measured data. For the 2C-process, the intensity scales quadratically with the pump intensity, leading to a calculated convolution as:

\begin{align}
    I_{2C}(\Delta t) \propto (I_{pr}^2 \ast I_s)(\Delta t).
\end{align}

Consequently, from theory we estimate a Gaussian shape with an FWHM of $\tau = \SI{2.26}{\pico\second}$, which explains the FWHM of the measured 2C-NRB of $\tau_{2C-NRB} = \SI[separate-uncertainty=true]{2.69 \pm 0.08}{\pico\second}$ to be narrower compared to the 3C-NRB duration. These calculations both underestimate the experimentally observed pulses due to the assumption of purely Gaussian pulses, while the temporal envelopes in the experiment are not ideally Gaussian-shaped. Nevertheless, the calculations explain the observed differences in decay times in the 2C and 3C-regimes.

For an experiment, this means that NRB-free and 2C-free spectra can be obtained in our setup for delays larger than approximately $\Delta t > \SI{3}{\ps}$.

\section{Phonon dephasing in diamond}

\begin{figure*}[htb]
	\centering
	\includegraphics[width=1\textwidth]{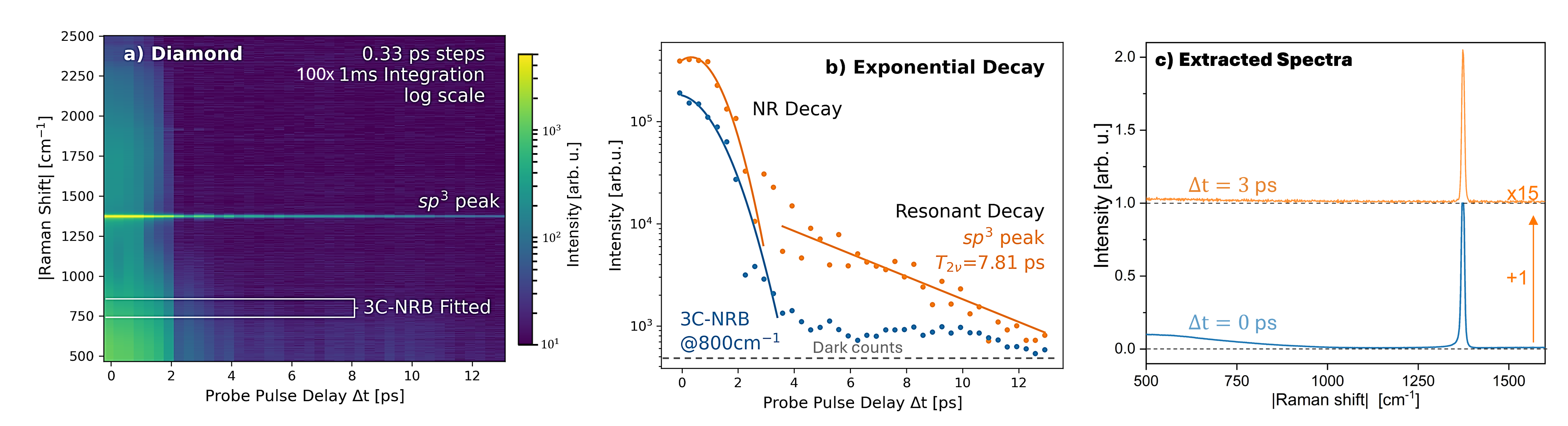}

    \caption[Time-Delay Sweep Diamond]{TD-BCARS spectra of diamond when varying the probe pulse delay: a) The 3C-NRB and 2C-NRB decay fast, while the strong $sp^3$ diamond peak is measurable even at \SI{10}{\pico\second} delay. b) On a logarithmic scale, both the 3C-NRB and the $sp^3$ signal exhibit a Gaussian-shaped decay region, while the resonant $sp^3$ signal shows an additional, linear decay. c) A symmetric, Lorentz-shaped peak form can be obtained with a time-delay of 3~ps.}
	\label{fig:TD_dia}
\end{figure*}

With the quantification of NRB-decay time being established, the behavior of the resonant signal can be investigated. Here, a single crystal of diamond in (100)-cut is chosen. The sample is a CVD-grown diamond commercially obtained from  Applied Diamond Inc., Wilmington (DE), USA. Diamond crystallizes in the eponymous diamond structure with two atoms per unit cell, which has a single three-times-degenerated optical phonon at the zone center. Hence, only a single peak belonging to this $sp^3$-bond is expected \cite{Solin1970}. Due to the strong bond and the relatively light C atom, its expected frequency is very high for a phonon in a typical crystal at around \SI{1336}{\per\cm}, which is at the edge of the 3C-frequency range. However, time-delayed measurements are still possible in this regime, which shows that our setup is capable of addressing most typical crystalline materials including many semi-conductors or oxides. Fig.~\ref{fig:TD_dia}~a) shows a hyperspectral map similar to Fig.~\ref{fig:TD_NRB}~a). Here, again the 2C-NRB and 3C-NRB regions are visible, but also an additional, resonant $sp^3$ peak at \SI{1336}{\per\cm}, which is measurable even for $\Delta t > \SI{10}{\pico\second}$. For time delays larger than 3~ps, its intensity is exclusively stemming from the 3C-process. For shorter time delays, this peak lays in the overlap regions of the 2C and 3C processes.


Figure~\ref{fig:TD_dia}~b) shows the intensity profiles in logarithmic scale as a function of time-delay for the $sp^3$ peak intensity and the 3C-NRB intensity around \SI{800}{\per\cm} in orange and blue, respectively. Here, as established for glass (see Fig.~\ref{fig:TD_NRB}) the NRB completely vanishes in the 3C-region within approximately 3~ps. In contrast, the intensity of the $sp^3$ peak first follows the behavior of the 3C-NRB (up to 3~ps of delay), but then continues as an exponential decay. This means that NRB-free spectra can be obtained for delays of $>3$~ps consistent with the plots in Fig. 3b), where larger delays result in even less NRB contribution, but at the cost of a reduced resonant peak intensity. The datapoints show a rather large scattering especially in the linear regime. This is not due to noise in the detector, but a result of the mechanical delay-stage, which causes slight random variations in coupling. These are amplified by the nonlinear nature of the process, but otherwise do not impact the fit.

For delays $\Delta t > \SI{4}{\pico\second}$, the $sp^3$ peak can be accurately fitted using an exponential decay function~\cite{Lee2008VibrationalDT}, which allows to determine the phononic dephasing time:

\begin{align}
    I(\Delta t) = I_0 e^{-2\Delta t / T_{2\nu}}.
\end{align}

\noindent Here, $I(\Delta t)$ describes the measured intensity as a function of time-delay $\Delta t$, and $I_0$ denotes the maximum intensity of this second decay region at $\Delta t = \SI{4}{\pico\second}$. The $T_{2\nu}$ describes the dephasing time of the vibrational coherence, which contains contributions from the vibrational population decay and the dephasing of the vibrational modes~\cite{Hamaguchi1994UltrafastTS}. Please note, the decay time $T_{decay}$ in the time-dependent spectra is not equal to the dephasing time $T_{2\nu}$. It holds $T_{decay} = T_{2\nu}/2$ ~\cite{Lee2008VibrationalDT}. 

The fitting of the $sp^3$ signal results in $T_{2\nu} = \SI[separate-uncertainty=true]{7.8 \pm 1.3}{\pico\second}$. 
This value falls within the reported range of diamond phonon lifetimes in the literature, which range from 5.7 to \SI{7.0}{\pico\second}~\cite{Jnior2020LifetimeAP, Lee2010ComparingPD, Nakamura2016SpectrallyRD}.
Our measured value of $T_{2\nu}$ falls in the higher range of previously measured values, indicating the high homogeneity in the CVD-grown crystals~\cite{Jnior2020LifetimeAP, Lee2010ComparingPD, Nakamura2016SpectrallyRD}.

When we inspect the data of diamond displayed in Fig. \ref{fig:TD_dia} more closely, we can make another key observation about the role of the NRB in the CARS process on the total intensity. If we extrapolate the exponential decay of the resonant signal to a time-delay of $\Delta t = 0$ ps we obtain an approximate resonant contribution of $2 \cdot 10^4$ arb.u. However, if we subtract the NRB intensity in a non-resonant region with the resonant intensity (blue vs. orange line at  $\Delta t = 0$~ps) we observe a difference of about $10^5$ arb.u., which is an order of magnitude larger than the extrapolation would suggest. This sounds surprising at first, however, is expected based on the CARS mechanism. The physical origin of the NRB is not that of a linear background, but rather the NRB acts as a self-heterodyne amplifier of the weak resonant Raman signal \cite{Vernuccio2024,Solin1970}. Therefore, while the NRB might distort the spectrum, i.e. hide spectral features, it also provides amplification. However, in time-delay CARS for delays larger than the NRB dephasing time, only the pure resonant nonlinear susceptibility is measured without amplification, whose intensity value is further reduced due to the dephasing of the vibrational population.

Here, for the study on diamond and the study on the NRB in glass (Fig. \ref{fig:TD_NRB}) we used an integration time of 1~ms per spectrum, which is at the limit of the integration time possible with the CCD-camera detector (0.8~ms). To be able to detect the phonon dephasing time to long time-delays, we further averaged over 100 spectra, which yields a total integration time of 100~ms per spectrum in our experiment. Depending on the desired signal-to-noise ratio, lower integration times are possible as evident from the spectrum in Fig. \ref{fig:TD_dia}c). Moreover, the diamond peak is already at the edge of the 2C-excitation spectrum envelope, which further reduces the available intensity. Nevertheless, 100 ms is still fast compared to typical integration times in spontaneous Raman scattering times.
Our findings show that time-delayed BCARS adds another tool for spectroscopy due to its unique properties for the analysis of crystalline materials in terms of phonon dephasing time and crystal quality, even though it might not be the best choice in all situations for ultra-high-speed imaging, due to the lower signal at higher delay.

\section{Phonon dephasing in KTP and KTA}

\begin{figure*}[htb]
	\centering
	\includegraphics[width=0.850\textwidth]{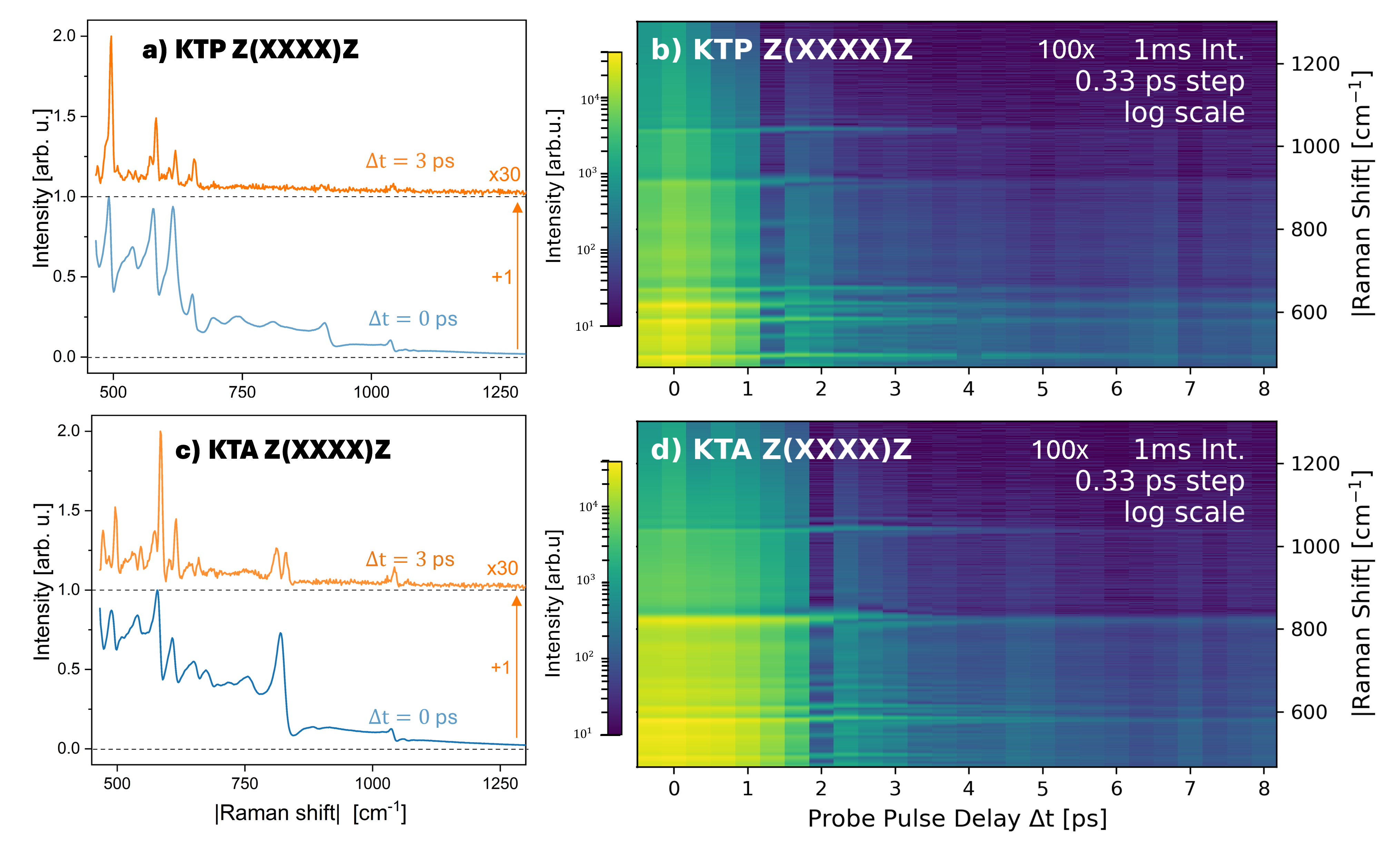}
	\caption[Time-Delay Sweep in KTP and KTA]{TD-CARS on Z(XXXX)Z KTP and KTA: a) The KTP spectrum at $\Delta t = \SI{0}{\ps}$ shows multiple distorted peaks, while with a delay of $\Delta t = \SI{3}{\ps}$ yields narrow peaks similar to spontaneous Raman scattering. b) The time-delay sweep shows the dephasing of the different peaks. The related material KTA is shown as a comparison in c) and d). Extracted phonon decoherence times are shown in Tab.~1.}
	\label{fig:TD_KTP}
\end{figure*}

To demonstrate the TD-BCARS methodology on more complex crystals, we investigated a sample of z-cut potassium titanyl phosphate (KTiOPO$_4$, KTP) and a sample of potassium titanyl arsenate (KTiOAsO$_4$, KTA), each. KTP and KTA are isostructural crystals, which see widespread use in nonlinear and quantum optics. From the viewpoint of Raman and CARS spectroscopy, KTP and KTA are ideal model systems as they feature a 64-atom unit cell with 189~optical phonons, which are all Raman- and CARS-active \cite{Ruesing2016,Neufeld2023,Bushiri1999,Watson1991,Kugel1988}. For one polarization setting of excitation and detection, typically more than 40 phonons are expected to be visible in one spectrum spanning a range from almost 0 to about 1200 cm$^{-1}$. The result of the broadband TD-BCARS measurement on KTP and KTA, each, is displayed in Fig.~\ref{fig:TD_KTP}.

Both KTP and KTA exhibit comparable Raman spectra, as shown in Figures~\ref{fig:TD_KTP}~a) and c). Interestingly, while the spectrum of KTP and KTA usually is expected to show more than 40 phonons per polarization setting, we do only see 10 to 15 peaks at a time-delay of 3~ps. Here, certain phonons may not be visible due to a fast dephasing. The hyperspectral plots in Figure~\ref{fig:TD_KTP}~b) and d) illustrate a range of resonant peaks as brighter lines against the NRB. While these peaks are not as prominent as the diamond signal, an averaging over 100 spectra with 1~ms integration time enabled acquisition for time-delays of up to \SI{8}{\ps}. Exponential decay fitting was applied to all peaks starting from $\Delta t = 3$~ps, and the resulting dephasing times $T_{2\nu}$ ranging from 2.7-\SI{7.6}{\pico\second} are presented in Table~\ref{tab:KTP_tau}. Here, distinct differences between vibrational modes in one material are visible indicating that the dephasing time is phonon dependent. In contrast, similar features in both materials feature comparable decay times indicating systematic similarities between both materials and their phonons.

The reported dephasing times of phonons in KTP from literature~\cite{Singhapurage2023DecayOR} are within the same order of magnitude. However, while the literature values range from 0.5-\SI{2.1}{\ps}, the measured results for our corresponding modes display lifetimes of 2.9-\SI{6.3}{\ps}. This hints at higher quality crystals in our experiment with longer phonon dephasing times. This experiment, as well as the previously discussed investigations on diamond suggest that dephasing times could be used to quantify the quality of crystals.

\begin{table*}[htbp]
\caption[KTP and KTA Dephasing Time]{Dephasing time $T_{2\nu}$ for phonons in KTP and KTA measured in Z(XXXX)Z configuration. The uncertainties in the peak positions are smaller than 2~cm$^{-1}$ and match well with known peaks from KTP and KTA.
\label{tab:KTP_tau}}
\centering
 \begin{tabular}{ |c|c|c|c|c|c| }
    \hline
    \hline
    Assigned & Decoherence & \multicolumn{2}{|c|}{KTP} & \multicolumn{2}{|c|}{KTA} \\
    crystal feature \cite{Ruesing2016,Neufeld2023,Bushiri1999,Watson1991,Kugel1988} & time~\cite{Singhapurage2023DecayOR} [ps]  & $\Delta\tilde{\nu}_{Peak}$ [cm$^{-1}$] & $T_{2\nu}$ [ps] & $\Delta\tilde{\nu}_{Peak}$ [cm$^{-1}$] & $T_{2\nu}$ [ps] \\
    \hline
    
    PO$_4$ / AsO$_4$ & & 469 &  6.2 $\pm$ 0.6 & 469 & 7.4 $\pm$ 1.7 \\ 
    PO$_4$ / AsO$_4$ & & 496 & 3.1 $\pm$ 0.3 & 495 & 4.2 $\pm$ 0.7  \\
    TiO$_6$        &      1.95 $\pm$ 0.05 & 544 &  5.7 $\pm$ 0.7 & 540 &  5.2 $\pm$ 1.1 \\
    TiO$_6$ & & 583 &  3.0 $\pm$ 0.2 & 580 &  2.8 $\pm$ 0.3 \\
    TiO$_6$ & & 614 & 3.7 $\pm$ 0.4 & 620 &  3.9 $\pm$ 0.5 \\
    TiO$_6$ & & 655 &  2.8 $\pm$ 0.2 & 657 &  5.0 $\pm$ 1.1  \\
    TiO$_6$ &  0.475 $\pm$ 0.004 & 680 & 6.1 $\pm$ 0.8 & 682 & 5.7 $\pm$ 1.3 \\
    TiO$_6$ & 1.64 $\pm$ 0.06 & 824 & 6.3 $\pm$ 1.0 & 815 &  7.3 $\pm$ 1.5  \\
    - && 917 & 6.5 $\pm$ 1.0  &917 &  7.6 $\pm$ 1.7  \\ 
    PO$_4$ / AsO$_4$ &2.15 $\pm$ 0.0& 1041 & 2.9 $\pm$ 0.3 & 1037 &  4.4 $\pm$ 0.4 \\
    \hline
    \hline
\end{tabular}
\end{table*}

\section{Conclusion}

Using  TD-BCARS, we demonstrate the simultaneous acquisition of NRB-free and broadband spectra in a wide spectral range in solid-state materials. Our experimental approach combines a broadband excitation with a time-delay mechanism and is designed to work in the typical vibrational frequency range of crystalline material from 0~cm$^{-1}$ up to about 1300~cm$^{-1}$. The methodology and experimental setup is tested on amorphous SiO$_2$ (glass), which is used to characterize the setup-specific and material-independent response times, and applied to single crystals of diamond and the ferroelectrics KTP and KTA. We measure dephasing times on the same order of magnitude as those acquired by complementary methods in single-frequency measurements, which renders the method a valuable extension for phononic examinations of materials. Recently, coherent excitations of phonons and their dephasing times see particular interest in experiments, where resonant, optical controls are used to switch material properties, like ferroelectric domains\cite{Chen2022DeterministicCO,Kwaaitaal2024EpsilonnearzeroRE}. Here, our fast measurement technique of full spectra could serve as a guide for such experiments. Another possible extension of our technique could be in imaging studies of solids using time delay as a contrast mechanism, which are expected to change depending on spatial variations of crystal structure or defects.

\section{Acknowledgements}
The authors gratefully acknowledge financial support by the Deutsche Forschungsgemeinschaft (DFG) through projects CRC1415 (ID: 417590517),
INST 269/656-1 FUGG and FOR5044 (ID: 426703838; \url{http:\www.For5044.de}), as well as the Würzburg-Dresden Cluster of Excellence ”ct.qmat” (EXC 2147; ID: 390858490), LASERLAB-EUROPE (grant agreement no. 871124, European Union’s Horizon 2020 research and innovation program), and the European Union project CRIMSON (grant agreement no. 101016923). GC and DP acknowledge financial support by the European Union’s NextGenerationEU Program with the I-PHOQS Infrastructure (IR0000016, ID D2B8D520, CUP B53C22001750006) “Integrated infrastructure initiative in Photonic and Quantum Sciences”. The authors thank Ji{\v{r}}{\'i} Hlinka of the Czech Academy of Sciences, Prague, Czech Republic, for providing samples.
\section{References}
\nocite{} 
\bibliography{aipsamp}

\end{document}